# Topological interface modes in local resonant acoustic systems

Degang Zhao[1,2], Meng Xiao[2], C. W. Ling[3], C. T. Chan[2], Kin Hung Fung[3,*]

[1]*Department of Physics, HuaZhong University of Science and Technology, Wuhan, 430074, China*

[2]*Department of Physics, Hong Kong University of Science and Technology, Clear Water Bay, Hong Kong, China*

[3]*Department of Applied Physics, The Hong Kong Polytechnic University, Hong Kong, China*

Topological phononic crystals (PCs) are periodic artificial structures which can support nontrivial acoustic topological bands, and their topological properties are linked to the existence of topological edge modes. Most previous studies focused on the topological edge modes in Bragg gaps which are induced by lattice scatterings. While local resonant gaps would be of great use in subwavelength control of acoustic waves, whether it is possible to achieve topological interface states in local resonant gaps is a question. In this article, we study the topological bands near local resonant gaps in a time-reversal symmetric acoustic systems and elaborate the evolution of band structure using a spring-mass model. Our acoustic structure can produce three band gaps in subwavelength region: one originates from local resonance of unit cell and the other two stem from band folding. It is found that the topological interface states can only exist in the band folding induced band gaps but never appear in the local resonant band gap. The numerical simulation perfectly agrees with theoretical results. Our study provides an approach of localizing the subwavelength acoustic wave.

## I. INTRODUCTION

Over the past decades, the topological properties have been attracting extensive research interests in electronic materials[1,2], mechanical metamaterials[3-5], photonic systems[6-8] and phononic systems[9-17]. Topological invariants, such as Chern number [1-17] and Zak phase[18-20], can be used to characterize the topological properties of bands. The most fascinating phenomenon of these topological systems is the bulk-edge correspondence associated with topologically protected edge modes, which are one-way propagation modes along the surface or interface and are immune to lattice defects. In one-dimensional (1D) periodic systems, the SSH model was firstly proposed by Su, Schrieffer and Heeger to explain the existence of edge modes in polyenes chain[21], and these states have been observed in experiment recently[22]. The SSH model has been served as a paradigmatic example of a 1D topological system and quickly extended to optical systems[23-26] and mechanical systems[27]. Recently, B. Zhang *et al*. proposed an acoustic SSH model consisting of 1D array of resonance cavities, and revealed topological interface states in lowest Bragg-like gap[28,29]. Most of the previous works focused on the topological properties in Bragg band gaps due to the Bragg scattering. In acoustic system, another essential mechanism to open band gaps is the local resonance, which stems from the resonance of individual scattering unit[33]. Can topological edge states exist in local resonant gaps? What are the topological invariants of the local resonant bands? To answer these questions, in this paper, we firstly build a spring-mass model to describe the bands of 1D SSH chain hybridized with extra local resonances and systematically demonstrate the evolution of band structure and topological properties in all three subwavelength band gaps. Finally, we verify the theoretical results in a practical 1D SSH lattice by numerical simulation.

*Email:khfung@polyu.edu.hk

## II. SPRING-MASS MODEL

Before we consider topological phononic systems with local resonances, we first begin with a simple infinitely long monatomic chain consisting of mass-in-mass local resonant unit cell, which is schematically presented in Fig.1(a). The unit cell can be modeled as a spherical shell of mass $M$ connected to a mass $m$ by two massless springs with equal spring constant $G$. All neighboring units are connected by massless springs with spring constant $K$ with a nearest center-to-center distance of $a$. The dispersion relation of such a mass-in-mass monatomic chain is given by

$$M_{eff}(\omega)\omega^2 = 4K\sin^2\frac{qL}{2}, \qquad (1)$$

where $M_{eff}(\omega) = M + \frac{m\omega_0^2}{\omega_0^2 - \omega^2}$ denotes the effective mass of the local resonant unit cell, $\omega_0 = \sqrt{2G/m}$ is the local resonance frequency, $q$ is the Bloch wave number and $L = a$ is the lattice constant. The dispersion relation in Eq. (1) is plotted in Fig.1(d) for $M = 2.0$, $m = 0.5$, $K = 1.25$, and $\omega_0 = 1.0$. A typical local resonant band gap appears when the frequency $\omega$ approaches $\omega_0$. At such local resonance frequency, we have $M_{eff} \to \infty$, meaning that the oscillator is effectively too heavy and does not have any response on external force. Consequently, a local resonant band gap is opened at $\omega \approx \omega_0$.

Next in Fig.1(b), we double the size of unit cell of that in Fig.1(a), i.e. $L = 2a$, and keep other parameters unchanged. The Brillouin zone for the new choice of unit cell in Fig.1(b) is thus reduced to a half of that in Fig.1(a). The band structure of this diatomic chain in Fig.1(b) can be obtained by the folding of band structure along the midline of Brillouin zone in Fig.1(d), which is shown in Fig.1(e). Further based on the structure of Fig.1(b), we make the oscillator's left and right springs have different spring constants $K_1$ and $K_2$ in Fig.1(c), which forms the so-called SSH chain[21]. Instead of writing a $4 \times 4$ matrix problem, we can transform the equations to an eigenvalue problem of $2 \times 2$ matrix:

$$\begin{pmatrix} 0 & K_1 + K_2 e^{-iqL} \\ K_1 + K_2 e^{iqL} & 0 \end{pmatrix} \begin{pmatrix} u_i \\ u_j \end{pmatrix} = \lambda \begin{pmatrix} u_i \\ u_j \end{pmatrix}, \qquad (2)$$

where $u_i$ and $u_j$ are the displacements of $M$ in one unit cell. The frequency dependence appears on the eigenvalues $\lambda(\omega) = [K_1 + K_2 - M_{eff}(\omega)\omega^2]$ and the eigenvectors only. In contrast to the standard mapping of either $\omega$ or $\omega^2$ to the eigen-energy, we have the eigenvalue $\lambda(\omega)$, which is a monotonic function in $\omega$ except the jump at $\omega_0$. Therefore, the eigenvalue problem in Eq. (2) forms two-to-one mapping to that of the SSH model. For example, the zero-energy eigenvalue in SSH model is thus mapped to the two positive frequencies $\omega_n$ satisfying the quartic (4th power) equation $K_1 + K_2 - M_{eff}(\omega_n)\omega_n^2 = 0$.

By searching for the non-trivial solutions of Eq. (2), we can obtain the following dispersion relation of SSH chain:

$$\cos qL = 1 - \frac{M_{eff}\omega^2(2K_1 + 2K_2 - M_{eff}\omega^2)}{2K_1 K_2}. \qquad (3)$$

The detailed derivation of Eqs. (1), (2) and (3) can be found in Appendix A, and the corresponding band structure calculated by Eq.(3) is plotted in Fig.1(f) with $K_1 = 1.0$, $K_2 = 1.5$ and all the other parameters are identical to that in Fig.1(e). As is well known, the patterns of local resonant band gap and Bragg band gap are very different. Generally speaking, for the local resonant band gap, its lower band terminates at the boundary of Brillouin zone and its upper band emerges at the center of Brillouin zone (see Fig.1(d)). But for the Bragg band gap, the termination of lower band and emergence of upper band will both locate at the boundary or center of Brillouin zone. By simply observing the patterns of band structures, all three band gaps in Fig.1(f) seem to be Bragg band gaps. But from the evolution of band structures (from Fig.1(d) to Fig.1(f)), it suggests that band gap II originates from the local resonance of individual local resonant unit. We also see that it survives in all three different structures. When $K_1 = K_2 = K$, the first band and second band, third band and forth band will meet at the folding points (see Fig.1(e)), which can be regarded as the topological phase transition point[19,28,29]. When $K_1 \neq K_2$, these two band crossing points are opened to be gaps I and III (see Appendix B), which can be considered as the band folding induced gaps.

We now investigate the topology of all bulk bands by evaluating the Zak phase[18]. Combining Eqs. (A3) and (A9) in Appendix A, we can obtain the eigenvector of SSH chain:

$$(u_i, v_i, u_j, v_j)^T = (\pm e^{i\phi(q)}, \pm \frac{\omega_0^2}{\omega_0^2 - \omega^2} e^{i\phi(q)}, 1, \frac{\omega_0^2}{\omega_0^2 - \omega^2})^T, \quad (4)$$

where $\phi(q) = \arg(K_1 + K_2 e^{-iqL})$. The superscript $T$ denotes the transpose operation. The $\pm$ signs in Eq. (4) are determined by the signs of eigenvalues of Eq. (2) (See Appendix C). The Zak phase[19,23] is thus obtained from the eigenvectors as

$$\theta = i\int_{-\pi/L}^{\pi/L} dq \langle \Psi | \partial_q | \Psi \rangle = -\frac{\phi(\frac{\pi}{L}) - \phi(-\frac{\pi}{L})}{2} = \begin{cases} \pi, & K_1 < K_2 \\ 0, & K_1 > K_2 \end{cases}. \quad (5)$$

where $|\Psi\rangle$ is the eigenvector normalized by the kinetic energy. The detailed calculation of Zak phase can be found in Appendix D. Eq. (5) denotes that all four bulk bands in Fig.1(f) have the same quantized Zak phase, either $0$ or $\pi$. The lattice in Fig.1(c) has only one definite band structure, but the different choice of unit cell ($K_1 < K_2$ or $K_1 > K_2$) would lead to different topological properties of bands. If two semi-infinite SSH chains with $K_1 < K_2$ and $K_1 > K_2$ are combined to be a whole block, the existence of topological edge states in the $n$th band gaps can be determined by the bulk-edge correspondence. We only need to know the sign of $\varsigma^{(n)}$

which is defined as[19]

$$\text{sgn}[\varsigma^{(n)}] = (-1)^n \exp(i \sum_{m=0}^{n-1} \theta^{(m)}) . \tag{6}$$

For the $n$th band gaps, if the $K_1 < K_2$ chain and $K_1 > K_2$ chain have different signs of $\varsigma^{(n)}$, topological edge states will definitely exist in the band gaps. Conversely if they have the same sign of $\varsigma^{(n)}$, topological edge states cannot exist. We calculate the sign of $\varsigma^{(n)}$ for all three band gaps and list them in table 1. From table 1, for two distinct SSH chains, $\varsigma^{(n)}$ have different signs only in band gaps I and III, but have the same signs in band gap II. That means the topological edge states can only exist in band gaps I and III, but not exist in band gap II. In other words the original local resonant band gap will not support any topological edge states, which are produced simultaneously in two band folding induced band gaps.

|  | $K_1 < K_2$ | $K_1 > K_2$ |
|---|---|---|
| $\text{sgn}[\varsigma^{(\text{I})}]$ | + | − |
| $\text{sgn}[\varsigma^{(\text{II})}]$ | + | + |
| $\text{sgn}[\varsigma^{(\text{III})}]$ | + | − |

Table 1. Sign of $\varsigma^{(n)}$ for diatomic chains with $K_1 < K_2$ and $K_1 > K_2$.

Another criteria to determine the existence of topological edge states is the switching of field distribution of eigenstates at the two band-edge points (marked by S1,S2,…, S6 in Fig.1(f)) of each isolated band gap for two different SSH chains, which is known as "band inversion" transition[2,19]. Since the two different SSH chains have identical band structure, we only need to investigate the sign of components $u_i, v_i, u_j, v_j$ in eigenvector, which are listed in table 2. The detailed calculation can be found in Appendix C. For the band folding induced gaps, viz. band gap I and III, the states of the lower band-edge and upper band-edge indeed switch when we change $K_1 < K_2$ to $K_1 > K_2$. But the states do not have any change in the band gap II, viz. the local resonance band gap. Then the topological edge states will definitely only exist in band gaps I and III, but not exist in band gap II. This result is consistent with the analysis of Zak phase.

|  |  | $K_1 < K_2$ | $K_1 > K_2$ |
|---|---|---|---|
| Band gap I | S1 | $(-,-,+,+)^T$ | $(+,+,+,+)^T$ |
|  | S2 | $(+,+,+,+)^T$ | $(-,-,+,+)^T$ |
| Band gap II | S3 | $(-,-,+,+)^T$ | $(-,-,+,+)^T$ |
|  | S4 | $(+,-,+,-)^T$ | $(+,-,+,-)^T$ |
| Band gap III | S5 | $(-,+,+,-)^T$ | $(+,-,+,-)^T$ |
|  | S6 | $(+,-,+,-)^T$ | $(-,+,+,-)^T$ |

Table 2. Sign of eigenstates at band-edges.

In order to verify the analytical derivations, we construct a supercell spring-mass system to numerically calculate the dispersion of interface states. The supercell is composed a SSH chain with $K_1 < K_2$ on the left-hand side connected to another SSH chain with $K_1 > K_2$ on the right-hand side. The band structures of connected chain is shown in Fig.2(a). It distinctly exhibits two extra flat bands, i.e. topological interface states appearing only in the first and third band gaps, which agree with our analytical prediction. To demonstrate the evolution of topological interface states, we keep the left-hand side chain unchanged, and investigate the dependence of band structure on the ratio of spring constants $K_1/K_2$ of right-hand side chain, which is exhibited in Fig.2(b). When $K_1/K_2 < 1$, all bulk bands of the left and right chains, although having different patterns, have the same topological properties. Based on the bulk-edge correspondence, the topological interface states certainly will not appear in common band gaps. But when $K_1/K_2 > 1$, the topology of all bulk bands of right diatomic chain has been converted, and obviously topological interface states emerge in the first and third band gaps. Thus these interface states are topologically protected and robust at the same frequency.

### III. SUBWAVELENGTH TOPOLOGICAL INTERFACE STATES IN A PRACTICAL SYSTEM.

In this section we propose a practical structure to achieve topological interface states in subwavelength region. The system is a 1D array of "core-shell" cylinders, which is presented in

Fig.3(a). Each cylinder consists of epoxy core (mass density $\rho = 1180\,\text{kg}/\text{m}^3$, longitudinal wave velocity $c_l = 2540\,\text{m}/\text{s}$ and transverse wave velocity $c_t = 1160\,\text{m}/\text{s}$) coated by soft rubber (mass density $\rho = 1300\,\text{kg}/\text{m}^3$, longitudinal wave velocity $c_l = 50\,\text{m}/\text{s}$ and transverse wave velocity $c_t = 20\,\text{m}/\text{s}$) and they are immerged in water background (mass density $\rho = 1000\,\text{kg}/\text{m}^3$, longitudinal wave velocity $c = 1490\,\text{m}/\text{s}$). We adjust the separation of two cylinders in one unit cell to mimic the different spring constant $K_1$ and $K_2$ in the spring-mass model. The separation of left lattice is $d_L(<L/2)$ and that of right lattice is $d_R(>L/2)$, and for simplicity we set $d_L + d_R = L$ to guarantee the left lattice and right lattice to have identical band structures. We perform the numerical simulation by finite-element commercial software COMSOL Multiphysics. The lowest four bulk bands of individual left or right lattice are plotted in Fig.3(b). All three band gaps are in subwavelength region with the normalized frequencies being lower than $0.08$. According to the analysis of spring-mass model, we can identify that the gap I and III are the band folding induced gaps, while gap II is the original local resonant band gap. We plot the pressure field distribution of eigenstate in one unit cell at the band-edges of all three gaps in Fig.3(c). For the left lattice, in the first and third gaps, the field distribution of the lower band-edges, *viz.* S1 and S5 are even modes since they are symmetric with respect to the central plane of unit cell. While the field distribution of the upper band-edges, *viz.* S2 and S6 are odd modes since they are antisymmetric with respect to the central plane of unit cell. However these symmetry properties are just reversed for the right lattice. So it is exactly that the band inversion happens in the first and third gaps. But for the second gap of both left and right lattice, the state at the lower band-edge S3 is an even mode, while that at the upper band-edge S4 is an odd mode. There is no band inversion in the second gap.

Next we calculate the transmission spectrum of a finite size lattice to examine the existence of topological interface states. The sample is composed by the connection of 8 periods left lattice with $d_L(<L/2)$ and 8 periods right lattice with $d_R(>L/2)$. From the transmission spectrum plotted in Fig.4(a), two topological interface states, manifesting as two sharp transmission peaks marked by A and B emerge in the first and third band gaps, respectively. And not any transmission peak exist in the second band gap. The pressure field distribution of these two peaks are plotted in Fig.4(b). Both of them exhibit the typical pattern of field distribution of interface states: the field intensity has the strongest value on the interface, and rapidly decays away from the surface.

## IV. CONCLUSION AND DISCUSSION

In this paper, we apply the spring-mass model to demonstrate the band evolution and topological properties of acoustic systems with local resonant unit cell. Owing to band folding, two Bragg-like gaps can be produced in subwavelength region. According to the calculation of Zak phase and analysis of band inversion, the topological interface states can only exist in the band folding induced gaps, but never exist in the original local resonant band gap. We further proposed a practical structure to achieve topological interface states in subwavelength region. Physically the local resonant band gap is due to the singularity of effective mass. When the frequency $\omega$ approaches to the resonant frequency $\omega_0$, the effective mass will tend to infinity and acoustic wave cannot propagate in it. In other words, near the resonant frequency, the band gap will definitely be opened. The lower and upper bands of local resonant band gap cannot meet each other to generate a topological phase transition point. Then it does not possess the essential condition to produce the "open-close-reopen", *viz.* band inversion process. Obviously the original local resonant band gap cannot support any topological interface states. However, it does not mean we cannot obtain topological interface states in subwavelength region. As shown in this paper, we can first produce pass bands in subwavelength frequency region, and then utilize band folding mechanism to open new gaps. Essentially these gaps are created by the different interactions between an oscillator and its left and right neighboring oscillators. It is easy to achieve band inversion by adjusting the separation of two oscillators in a unit cell and finally obtain topological interface states in these gaps. Actually similar band folding mechanism has been applied in the Bragg gaps in two-dimensional systems by other authors[17,34]. The method proposed in this paper can be treated as a general methodology to obtain topological edge states in subwavelength region. It can be smoothly expanded to negative effective modulus system, such as Helmholtz resonator chain[35], as well as two-dimensional and three-dimensional optical, mechanical, acoustic or elastic systems.

## ACKNOWLEDGMENTS

We thank Zhaoqing Zhang, Kun Ding, Shubo Wang and Wenjie Chen for helpful discussions. This work is supported by Research Grants Council, University Grants Committee, Hong Kong (Grant No. AoE/P-02/12).

**APPENDIX A: SPRING-MASS MODEL**

For the monatomic chain schematically presented in Fig.1(a), the equations of motion for the $n$th unit cell are:

$$M \frac{d^2 u_n}{dt^2} = K(u_{n-1} + u_{n+1} - 2u_n) + 2G(v_n - u_n), \tag{A1}$$

$$m \frac{d^2 v_n}{dt^2} = 2G(u_n - v_n), \tag{A2}$$

where $u_n$ and $v_n$ denote the displacement of mass $M$ and $m$ with respect to their equilibrium position in the unit cell, respectively. The displacement of an infinite atomic chain has harmonic solution $u(v) = A_{u(v)} e^{i(kx - \omega t)}$, where $A$, $k$ and $\omega$ are the amplitude, wave number and angular frequency, respectively. By substituting the harmonic solution, Eq.(A2) becomes

$$v_n = \frac{2G}{2G - m\omega^2} u_n, \tag{A3}$$

Substituting Eq.(A3) into Eq.(A1) to eliminate $v_n$, we obtain

$$-M_{eff}\omega^2 u_n = K(u_{n-1} + u_{n+1} - 2u_n), \tag{A4}$$

where

$$M_{eff} = M + \frac{m\omega_0^2}{\omega_0^2 - \omega^2}, \quad \omega_0^2 = \frac{2G}{m}. \tag{A5}$$

Eq. (A4) is the standard vibration equation of a periodic monatomic chain, and Eqs. (A4) and (A5) indicate that local resonant unit cell can be regarded as an effective single oscillator with an effective mass $M_{eff}$ [30-32]. Applying the Bloch's theorem $u_{n\pm 1} = e^{\pm iqL} u_n$ ($q$ is the Bloch wave number, $L = a$ is the lattice constant) in Eq. (A4), the dispersion relation of monatomic chain can be obtained as

$$M_{eff}\omega^2 = 4K \sin^2 \frac{qL}{2}. \tag{A6}$$

For the diatomic chain schematically presented in Fig.1(c), the equations of motion become

$$-M_{eff}\omega^2 u_i = K_1(u_j - u_i) + K_2(u_{j-1} - u_i), \tag{A7}$$

$$-M_{eff}\omega^2 u_j = K_2(u_{i+1} - u_j) + K_1(u_i - u_j). \tag{A8}$$

Applying the Bloch's theorem $u_{i+1} = e^{iqL} u_i$, $u_{j-1} = e^{-iqL} u_j$, the above equations can be written as matrix form

$$\begin{pmatrix} 0 & K_1 + K_2 e^{-iqL} \\ K_1 + K_2 e^{iqL} & 0 \end{pmatrix} \begin{pmatrix} u_i \\ u_j \end{pmatrix} = (K_1 + K_2 - M_{eff}\omega^2) \begin{pmatrix} u_i \\ u_j \end{pmatrix}. \tag{A9}$$

That's the eigenvalue equation of diatomic chain

**APPENDIX B: DISPERSION RELATION SOLUTIONS AT THE BAND-EDGES**

For the dispersion relation Eq. (3) in main text, we let $x = M_{eff}\omega^2$. The equation is simplified to be

$$\cos qL = 1 - \frac{x(2K_1 + 2K_2 - x)}{2K_1 K_2}. \tag{B1}$$

For the pass bands, Bloch wave number $q$ should be real. Then we obtain inequation

$$\left|1 - \frac{x(2K_1 + 2K_2 - x)}{2K_1 K_2}\right| \leq 1. \tag{B2}$$

Without loss of generality, we assume $K_1 < K_2$, the solution of Eq. (B2) is

$$0 \leq x \leq 2K_1 \quad or \quad 2K_2 \leq x \leq 2K_1 + 2K_2. \tag{B3}$$

At the band-edges, i.e. $q = 0$ or $q = \pi/L$, we can obtain the exact solutions of $x$ (use equality sign in Eq.(B3)) and mark them in Fig. B1. It clearly reveals that the first and third band gaps are opened because of $K_1 \neq K_2$. And the second band gap is opened due to the singularity of $M_{eff}$, since $\omega_0 = 1.0$ is the central frequency of it. When $K_1 = K_2 = K$, $x$ has a double root $2K$ at $q = \pi/L$, which corresponds to the band crossing points.

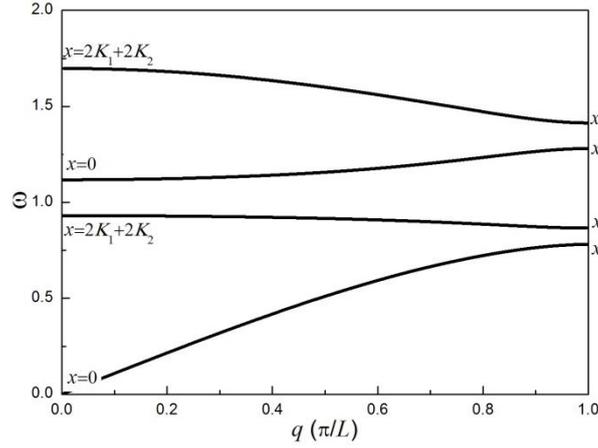

Fig.B1. Solutions of $x$ in Eq. (B1) at all band-edges.

## APPENDIX C: SIGNS OF EIGENVECTORS AT THE BAND-EDGES

The eigenvalue of Eq. (2) in main text can be written as

$$\lambda = K_1 + K_2 - M_{eff}\omega^2 = \pm\left|K_1 + K_2 e^{-iqL}\right|. \tag{C1}$$

The sign in eigenvector in Eq. (4) in main text is equivalent to the sign of $\lambda$, which is determined by the sign of $K_1 + K_2 - M_{eff}\omega^2$. We set function

$$\lambda(\omega) = 2K_1 + 2K_2 - M_{eff}\omega^2 = 2K_1 + 2K_2 - (M + \frac{m\omega_0^2}{\omega_0^2 - \omega^2})\omega^2. \tag{C2}$$

Then

$$\frac{d\lambda}{d\omega} = -2\omega M - \frac{2m\omega\omega_0^4}{(\omega_0^2 - \omega^2)^2} < 0. \tag{C3}$$

It denotes that function $\lambda(\omega)$ is a monotonic decreasing function except the jump at $\omega_0$. Then all four bulk bands can be divided into two sections: the first and second bulk bands are in the $\omega < \omega_0$ region while the third and fourth bulk bands are in the $\omega > \omega_0$ region. Since $\lambda(\omega)$ is a decreasing function, in each region, the lower (first and third) bands have positive eigenvalues and the upper (second and fourth) bands have negative eigenvalues, which is clearly present in Fig. C1. Then we can calculate the sign of eigenstates at all band-edges for $K_1 < K_2$ and $K_1 > K_2$ cases. For example, when $K_1 < K_2$, at state S1 (marked in Fig.1(c)),

$$\phi(q) = \arg(K_1 + K_2 e^{-iqL})\big|_{qL=\pi} = \pi. \tag{C4}$$

Substituting Eq. (C4) into Eq. (4) in main text, and noticing that the sign before the first two components in eigenvector are positive since $\lambda > 0$,

$$\operatorname{sgn}(u_i, v_i, u_j, v_j)^T = \operatorname{sgn}(e^{i\phi(q)}, \frac{\omega_0^2}{\omega_0^2 - \omega^2} e^{i\phi(q)}, 1, \frac{\omega_0^2}{\omega_0^2 - \omega^2})^T = (-,-,+,+)^T. \tag{C5}$$

While at state S2, $\phi(q)$ still equals to $\pi$ since it located on the Brillouin zone boundary too. But now $\lambda < 0$, the sign of eigenvector becomes

$$\operatorname{sgn}(u_i, v_i, u_j, v_j)^T = \operatorname{sgn}(-e^{i\phi(q)}, -\frac{\omega_0^2}{\omega_0^2 - \omega^2} e^{i\phi(q)}, 1, \frac{\omega_0^2}{\omega_0^2 - \omega^2})^T = (+,+,+,+)^T. \tag{C6}$$

When $K_1 > K_2$, at both states S1 and S2,

$$\phi(q) = \arg(K_1 + K_2 e^{-iqL})\big|_{qL=\pi} = 0. \tag{C7}$$

Sign of eigenvector at S1 is

$$\operatorname{sgn}(u_i, v_i, u_j, v_j)^T = \operatorname{sgn}(e^{i\phi(q)}, \frac{\omega_0^2}{\omega_0^2 - \omega^2} e^{i\phi(q)}, 1, \frac{\omega_0^2}{\omega_0^2 - \omega^2})^T = (+,+,+,+)^T. \tag{C8}$$

Sign of eigenvector at S2 is

$$\operatorname{sgn}(u_i, v_i, u_j, v_j)^T = \operatorname{sgn}(-e^{i\phi(q)}, -\frac{\omega_0^2}{\omega_0^2 - \omega^2} e^{i\phi(q)}, 1, \frac{\omega_0^2}{\omega_0^2 - \omega^2})^T = (-,-,+,+)^T. \tag{C9}$$

Eqs. (C5), (C6), (C8) and (C9) distinctly presents a band inversion. The signs of eigenstates at all the other band-edges can be obtained similarly.

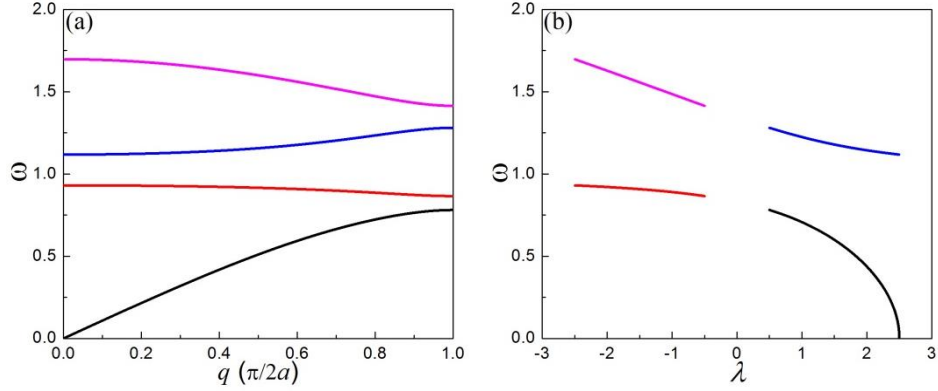

Fig.C1. (a) Band structure of SSH chain. All bands are distinguished by different colors. (b) Eigenvalues for all bands. The colors are one-to-one correspondence to that in (a).

## APPENDIX D: CALCULATION OF ZAK PHASE

Since our eigenvalue problem (Eq.(2) in main text) is not written as Hamiltonian form, the eigenvector (Eq.(4) in main text) cannot be directly used in calculating Zak phase. The eigenvector should be normalized by energy, and then it will definitely correspond to a Hamiltonian, even though we don't know its exact expression. At first we transfer the displacement to velocity

$$\frac{d}{dt}(u_i, v_i, u_j, v_j)^T = (\pm i\omega e^{i\phi(q)}, \pm i\omega \frac{\omega_0^2}{\omega_0^2 - \omega^2} e^{i\phi(q)}, -i\omega, -i\omega \frac{\omega_0^2}{\omega_0^2 - \omega^2})^T. \quad (D1)$$

It should be noted that $\omega$ is also a function of $q$. Normalize the eigenvector as

$$|\Psi\rangle = \frac{1}{\sqrt{M + m\left(\frac{\omega_0^2}{\omega_0^2 - \omega^2}\right)^2}} (\pm\sqrt{\frac{M}{2}}e^{i\phi(q)}, \pm\sqrt{\frac{m}{2}}\frac{\omega_0^2}{\omega_0^2 - \omega^2}e^{i\phi(q)}, -\sqrt{\frac{M}{2}}, -\sqrt{\frac{m}{2}}\frac{\omega_0^2}{\omega_0^2 - \omega^2})^T$$
$$= (\pm A(q)e^{i\phi(q)}, \pm B(q)e^{i\phi(q)}, A(q), B(q))^T$$

(D2)

where $2(A^2 + B^2) = 1$. Then the inner product of the eigenvector with itself gives the kinetic energy. Substituting $|\Psi\rangle$ in Eq.(5) in main text,

$$\theta = i\int_{-\pi/L}^{\pi/L}[\left(\pm A(q)e^{i\phi(q)}\right)^{*}\frac{\partial\left(\pm A(q)e^{i\phi(q)}\right)}{\partial q}+\left(\pm B(q)e^{i\phi(q)}\right)^{*}\frac{\partial\left(\pm B(q)e^{i\phi(q)}\right)}{\partial q}+A(q)\frac{\partial A(q)}{\partial q}+B(q)\frac{\partial B(q)}{\partial q}]\mathrm{d}q$$

$$= i\int_{-\pi/L}^{\pi/L}[2A(q)\frac{\partial A(q)}{\partial q}+iA^{2}(q)\frac{\partial\phi(q)}{\partial q}+2B(q)\frac{\partial B(q)}{\partial q}+iB^{2}(q)\frac{\partial\phi(q)}{\partial q}]\mathrm{d}q$$

$$= i\int_{-\pi/L}^{\pi/L}[\frac{\partial\left(A^{2}(q)+B^{2}(q)\right)}{\partial q}+i\left(A^{2}(q)+B^{2}(q)\right)\frac{\partial\phi(q)}{\partial q}]\mathrm{d}q$$

$$= -\frac{1}{2}\int_{-\pi/L}^{\pi/L}\frac{\partial\phi(q)}{\partial q}\mathrm{d}q = -\frac{1}{2}\left[\phi\left(\frac{\pi}{L}\right)-\phi\left(-\frac{\pi}{L}\right)\right]$$

(D3)

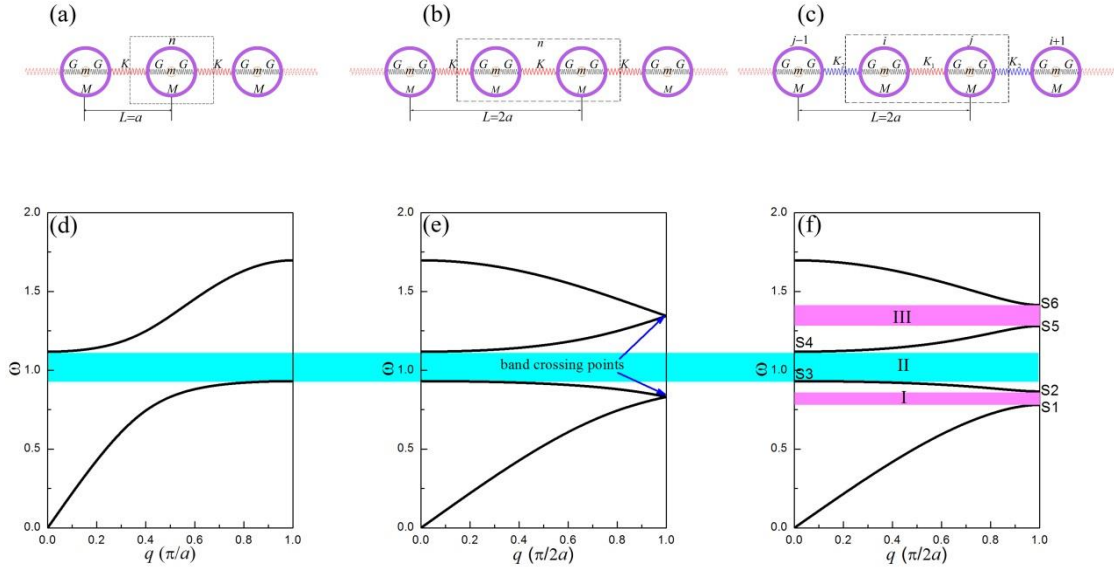

Fig.1. The schematic spring-mass model of 1D array of local resonant unit cell of (a): monatomic chain, (b): same as (a) except that the new unit cell consists of two unit cells in (a), and (c): diatomic chain with $K_1 \neq K_2$. The frames marked by dashed line are the unit cells for these three distinct chains. $L$ denotes the lattice constant. (d), (e) and (f) are the band structures for system (a), (b) and (c), respectively. The cyan and magenta strips represent the local resonant gaps and band folding induced gaps, respectively. Three gaps in (f) are numbered by Roman numbers I, II, and III. S1 to S6 in (f) denote the band-edges of all gaps.

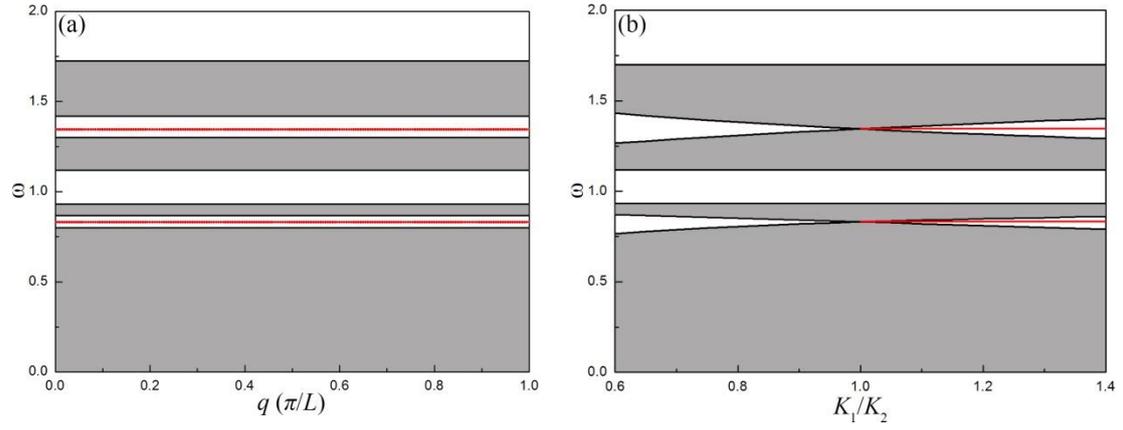

Fig.2. (a) Topological interface states (red lines) of a connected chain system, which is composed by SSH chain with $K_1 < K_2$ on the left-hand side and SSH chain with $K_1 > K_2$ on the right-hand side. The periods of left-hand side and right-hand side chains are 20. (b) Dependence of interface states (red lines) at $q = 0$ on $K_1/K_2$ of the right-hand side SSH chain, while the left-hand side SSH chain ($K_1 < K_2$) remain unchanged. For simplicity, we keep $K_1 + K_2$ constant in calculation. The gray shaded regions in both (a) and (b) denote the projection bulk bands.

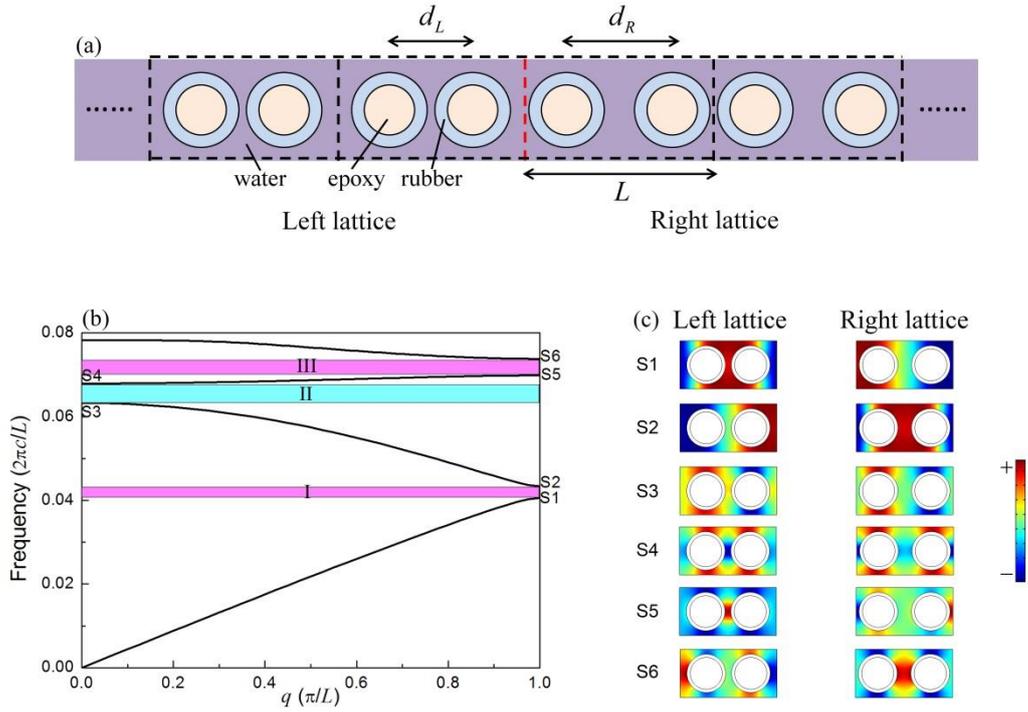

Fig.3. (a) Configuration of a practical structure which modelling the connected SSH chain with local resonant unit cell. Each scatterer is a cylinder which is composed by an epoxy core enwrapped by soft rubber, and all cylinders are put in water background. The dashed black frame denotes the unit cell. Separation of two individual cylinders in a unit cell is $d_L$ for the left lattice and $d_R$ for the right lattice. $L$ is the lattice constant. The inner and outer radii of cylinder are $0.165L$ and $0.2L$. The red dashed line denotes the interface between two lattices. (b) Band structure of left or right lattice. The frequencies are in unit of $2\pi c/L$, where $c$ is the acoustic wave velocity of water background. All symbols have the same meaning as that in Fig.1(f). (c) Pressure field distribution of band-edges for left lattice and right lattice, where only the pressure fields in the water background are shown.

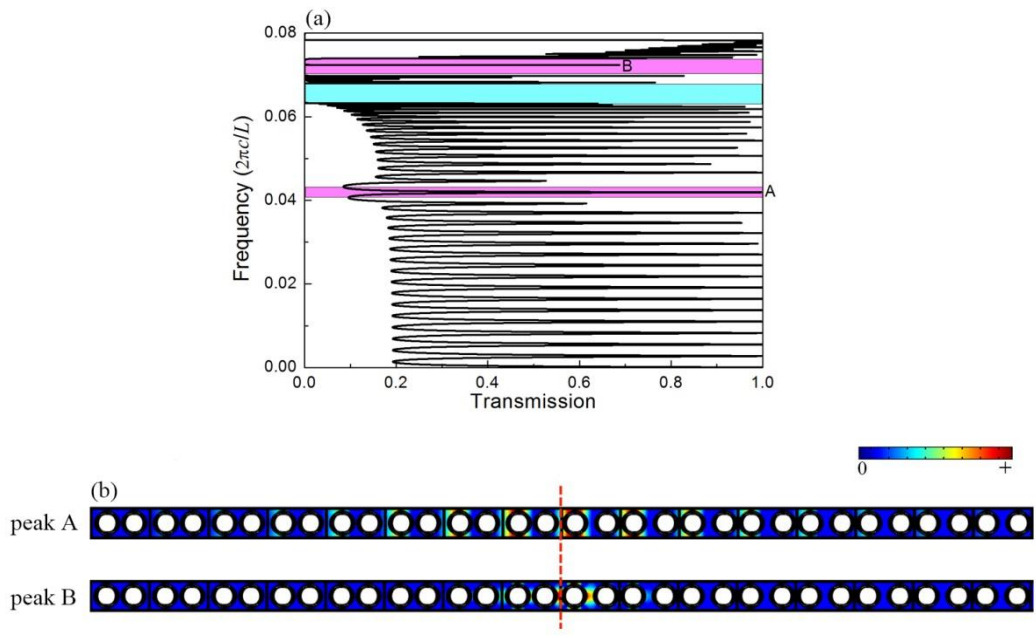

Fig.4. (a) Transmission spectrum of a finite size lattice which is shown in Fig.3(a). The number of period of left and right lattice is 8. Label A and B denote the transmission peaks in band gaps. (b) Absolute value of pressure field distribution (only inside the background) of peaks A and B.